\documentclass[12pt,aps,print,showpacs]{revtex4-1}
\usepackage{amsfonts}
\usepackage{amsmath}
\usepackage{amssymb}
\usepackage{amsthm}
\usepackage{graphicx}
\usepackage{subfigure}
\usepackage{hyperref}
\usepackage{xcolor}
\usepackage{cleveref}
\usepackage[matrix,frame,arrow]{xypic}

\input{Qcircuit}

\begin{document}

\title{Efficient quantum circuits for Toeplitz and Hankel matrices}
\author{A. Mahasinghe$^{1,2}$}
\author{J. B. Wang$^{2}$}
\email{jingbo.wang@uwa.edu.au}
\affiliation{{$^{1}$Department of Mathematics, University of Colombo, Colombo, Sri Lanka}}
\affiliation{{$^{2}$School of Physics, The University of Western Australia, Crawley WA 6009, Australia}}

\begin{abstract}
Toeplitz and Hankel matrices have been a subject of intense interest in a wide range of science and engineering related applications. In this paper, we show that quantum circuits can efficiently implement sparse or Fourier-sparse Toeplitz and Hankel matrices.  This provides an essential ingredient for solving many physical problems with Toeplitz or Hankel symmetry in the quantum setting with deterministic queries.
\end{abstract}

\maketitle
\section{Introduction}

Constructing an efficient quantum circuit to implement a given matrix operation is of fundamental importance in the field of quantum computation and quantum information.
One direct implication of the works by Aharanov, Ta-Shma, Childs and Berry \cite{aharanov2003, berry2014a,berry2007efficient,berry2007quantum} is that, the action of the exponent $e^{iH}$ of an arbitrary sparse Hermitian matrix $H$ can be efficiently implemented on a quantum state $\ket{\psi}$. An $n \times n$ matrix is row-sparse, if each row has at most $O($polylog$(n)$) nonzero entries. It is row-computable, if the non-zero elements in each row can be computed in runtime $O($polylog$(n)$). The sparse Hamiltonian lemma states that, if a Hermitian matrix $H$ is row-sparse, row-computable and $\left \| H \right \| \leq$  $O($polylog$(n)$), then $H$ is simulatable \cite{aharanov2003, berry2014a}. This means that the unitary operation $U=e^{-iH}$ can be approximated to an arbitrary accuracy $\epsilon$, using $O($polylog$(n)$, $\frac{1}{\epsilon}$) quantum gates. The sparse Hamiltonian lemma has been the key ingredient in a number of practically significant quantum algorithms~\cite{harrow2009,berry2014exponential}. Simplifying and generalising a number of previous quantum algorithms, Jordan and Wocjan \cite{jordan2009} proved that, if $U$ is unitary and sparse, it is possible to efficiently implement $U \ket{\psi}$ directly.  For an arbitrary diagonal unitary matrix, explicit quantum circuits can be found in \cite{childs_quantum_2004, welch2014}.

Broadening the quantum circuit framework to perform non-unitary operations is also of vital importance in order to extend the range of practically useful quantum algorithms. A quantum algorithm for the efficient implementation of $A^{-1}$ for an arbitrary sparse matrix $A$ was introduced by Harrow, Hassidim and Lloyd~\cite{harrow2009}, which is known as the HHL algorithm. Given a row-sparse matrix $A$, the HHL algorithm can implement $A^{-1} \ket{\psi}$ with runtime and resource cost $O($polylog$(n))$. Quantum circuit implementations and experimental verifications of the HHL methods can be found in \cite{cai2013, pub1257, pan2014}. The HHL algorithm relies upon the quantum Fourier transform, quantum eigenvalue estimation, and post-selection. Apart from its direct application in solving sparse linear systems, the HHL algorithm has been applied to solve several other important problems efficiently. A few examples are the $d$-dimensional Poisson equation \cite{cao2013}, systems of differential equations \cite{berry2014}, curve fitting and computing the effective resistance in electric networks  \cite{wang2014}. The HHL algorithm is thus an example of  high applicability of non-unitary operations performed through quantum circuits.

One implication of the HHL algorithm is that, quantum circuits can efficiently implement any sparse matrix as well as its inverse on any given quantum state.  Thus, it is natural to ask if any other matrices can be implemented efficiently through quantum circuits. It is well-known that implementing an arbitrary non-sparse matrix in a quantum circuit is a daunting task. This motivates us to investigate classes of efficiently implementable matrices, which also have a significant practical importance.

Two specific classes of matrices with a wide range of applications are the Toeplitz and Hankel matrices.  These matrices arise in different fields of physics, mathematics and engineering; such as quantum mechanics, signal processing, partial differential equations, differential geometry, and numerical integration \cite{ye2015, noschese2013, michael1979}.  Application of the Toeplitz and Hankel operators on states plays a key role in extracting information of systems with corresponding symmetries. A number of physical systems are analysed through structured Toeplitz and Hankel matrices~\cite{Steimel1979, Eisenberg1995, Serra1999, Cochran2001, Rietsch2003, Haupt2010}.  Moreover, any arbitrary matrix can be decomposed into a product of these matrices \cite{ye2015}.

A $n \times n$ Toeplitz (or Hankel) matrix can be fully described by an array of $2n-1$ elements. Using this fact, for a given $n \times n$ Toeplitz matrix, it is possible to define a unique array with $2n$ elements. The given Toeplitz matrix is row-sparse if and only if this array is sparse. Therefore, a $n \times n$ row-sparse Toeplitz matrix is fully characterised by an array of $2n$ elements, $O($polylog$(n))$ of them are non-zero. On the other hand, if the Fourier transform of this $2n$-element array has only $O($polylog$(n))$ non-zero elements (that is, if this array is sparse in the frequency domain instead of the time domain), we say the relevant Toeplitz matrix is \textit{frequency-sparse}. It is noteworthy that a frequency-sparse Toeplitz can be a non-sparse matrix.

We present an efficient quantum algorithm and circuit implementation for sparse and frequency-sparse Toeplitz (and Hankel) matrices. Our algorithm has several straightforward applications, such as  implementing Laplacians, solving circulant systems. This approach yields exponential speedups over existing classical algorithms. In the preceding section, we describe our algorithm for sparse and frequency-sparse Toeplitz matrices, and in section III we show that a slightly modified version would implement sparse and frequency sparse Hankel matrices efficiently.

\newpage

\section{Quantum algorithm}
A Toeplitz matrix is a matrix in which its elements are constant along all diagonals parallel to the main diagonal. More precisely, a Toeplitz matrix is a matrix of the form,
\begin{equation}
	T = \left( \begin{array}{ccccccc}
                    t_{0} & t_{-1} & t_{-2} & \cdots & t_{-\left( n-3 \right)} & t_{-\left( n-2 \right)} & t_{-\left( n-1 \right)} \\
                    t_{1} & t_{0} & t_{-1} & \cdots & t_{-\left( n-4 \right)} & t_{-\left( n-3 \right)} & t_{-\left( n-2 \right)} \\
		    t_{2} & t_{1} & t_{0} & \cdots & t_{-\left( n-5 \right)} & t_{-\left( n-4 \right)} & t_{-\left( n-3 \right)} \\
		    \vdots & \vdots & \vdots & \ddots & \vdots & \vdots & \vdots \\
		    t_{n-3} & t_{n-4} & t_{n-5} & \cdots & t_{0} & t_{-1} & t_{-2} \\
		    t_{n-2} & t_{n-3} & t_{n-4} & \cdots & t_{1} & t_{0} & t_{-1} \\
		    t_{n-1} & t_{n-2} & t_{n-3} & \cdots & t_{2 } & t_{1} & t_{0}
                  \end{array}
                \right).
\end{equation}

Thus, a $n \times n$ Toeplitz matrix is fully described by the $2n-1$ entries of its first row and column. With an extra zero element, we may associate the following array to the Toeplitz matrix.
\begin{equation}
\psi_T = \left( t_{0}, t_{-1}, \cdots, t_{-( n-1)}, 0, t_{n-1}, t_{n-2} \cdots t_{3}, t_{2}, t_{1} \right).
\end{equation}
An immediate observation is that, a Toeplitz matrix $T$ is row-sparse if and only if $\psi_T$ is sparse. Considering the rows as sequences of data corresponding to different moments of time, this can be described as the sparsity of $\psi_T$ in the \textit{time} domain. Similarly, we can also regard the sparsity of an array in the \textit{frequency} domain. Thus, we define a class of Toeplitz matrices as follows: a Toeplitz matrix $T$ so that $\psi_T$ is sparse in the frequency domain is called \textit{frequency-sparse}. More precisely, if the Fourier transform of $\psi_T$ has at most $O($polylog$(n))$ non-zero elements, the corresponding Toeplitz matrix $T$ is said to be frequency-sparse.

A special category of Toeplitz matrices with interesting spectral properties is the class of circulants -- square matrices in which the elements of each row are identical to those of the previous row, but are moved one position to the right and wrapped around \cite{davis1979}. The circulant $C = circ(c_{1}, c_{2}, ... , c_{n})$ is given by,
\begin{align*}
	C = \left( \begin{array}{cccc}
                    c_{1} & c_{2} & \cdots & c_{n} \\
                    c_{n} & c_{1} & \cdots & c_{n-1} \\
                    \vdots & \vdots & \ddots & \vdots \\
                    c_{2} & c_{3} & \cdots & c_{1} \\
                  \end{array}
                \right).
\end{align*}
One of the major advantages of using a circulant in quantum circuits is its diagonalisation. It is well known  that an $n \times n$ circulant is diagonalised by the discrete Fourier transform (DFT) matrix $F_{n}$,  i.e.\begin{equation}
C = F^{\dagger}_{n} diag\left( \lambda_{1}, \lambda_{2}, \dots, \lambda_{n} \right) F_{n},
\end{equation}
where $\lambda_{j}$ is the $j$-th eigenvalue of $C$, given by $\lambda_{j}=c_{1} + c_{2}\omega^{j-1} + c_{3}\omega^{(j-1)2} + \dots + c_{n}\omega^{(j-1)(n-1)}$, and $\omega=\exp{(2\pi i/n)}$.

Although a Toeplitz matrix is not circulant in general, any Toeplitz matrix $T$ can be embedded in a circulant defined by~\cite{michael1979},
\begin{equation}
C_{T} = \begin{pmatrix} T & B_T \\ B_T & T \end{pmatrix},
\end{equation}
where $B_T$ is given below
\begin{equation}
	B_T = \left( \begin{array}{ccccccc}
                    0 & t_{n-1} & t_{n-2} & \cdots & t_{3} & t_{2} & t_{1} \\
                    t_{-\left( n-1 \right)} & 0 & t_{n-1} & \cdots & t_{4} & t_{3} & t_{2}\\
		    t_{-\left( n-2 \right)} & t_{-\left( n-1 \right)} & 0 & \cdots & t_{5} & t_{4} & t_{3} \\
		    \vdots & \vdots & \vdots & \ddots & \vdots & \vdots & \vdots \\
		    t_{-3} & t_{-4} & t_{-5} & \cdots & t_{0} & t_{n-1} & t_{n-2} \\
		    t_{-2} & t_{-3} & t_{-4} & \cdots & t_{-\left( n-1 \right)} & t_{0} & t_{n-1} \\
		    t_{-1} & t_{-2} & t_{-3} & \cdots & t_{-\left( n-2 \right)} & t_{-\left( n-1 \right)} & t_{0} \\
                  \end{array}
                \right).
\end{equation}

Our objective is to implement the operation of $T $ on a quantum state vector $\ket{\psi}$ efficiently, using a quantum circuit, for any given row-sparse or frequency-sparse Toeplitz matrix $T$. It can be readily seen that,
\begin{equation}
\begin{pmatrix} T & B_T \\ B_T & T \end{pmatrix}
\begin{pmatrix} \psi \\ 0 \end{pmatrix}
=
\begin{pmatrix} T \psi \\ B_{T} \psi \end{pmatrix}.
\end{equation}
Considering the diagonalisation of a circulant, we observe that both $F$ and $F^{\dagger}$ are unitary operations; however, the diagonal matrix consisting of the eigenvalues may not be unitary in general. In order to embed it in a unitary matrix, we make use of the unitary dilation. Denoting the diagonal matrix $diag\left( \lambda_{1}, \lambda_{2}, \dots, \lambda_{n} \right)$ by $\Lambda_{C_{T}}$, we have the unitary dilation $U(\Lambda_{C_{T}})$ of ${\Lambda_{C_{T}}}$ given by,
\begin{equation}
U(\Lambda_{C_{T}}) =
 \begin{pmatrix}
  \frac{1}{k} \Lambda_{C_{T}} & \sqrt{I- \frac{1}{k^2}  \Lambda_{C_{T}} \Lambda_{C_{T}}^{\dagger}} \\
  \sqrt{I- \frac{1}{k^2} \Lambda_{C_{T}}^{\dagger} \Lambda_{C_{T}} } & -\frac{1}{k} \Lambda_{C_{T}}^{\dagger} \\
 \end{pmatrix},
\end{equation}
where $k$ is the square-root of the maximum modulus of $C_{T}$, i.e.
\begin{equation}
k=\sqrt{\left( max \{ \left| \lambda_j \right|: j=1,2,\cdots, n \} \right)}.
\end{equation}

If our chosen Toeplitz $T$ is row-sparse, the circulant $C_{T}$ is row-sparse as well.  Since each of its eigenvalues is a function of $t_{j}$'s and $\omega$ and we have polynomially many $t_{j}$'s in $T$, each eigenvalue is also efficiently computable. That is, the two non-zero elements in each row of $U(\Lambda_{C_{T}})$ are efficiently computable, which proves that $U(\Lambda_{C_{T}})$ is a row-computable, 2-sparse unitary matrix.

If $T$ is frequency-sparse, then the array $\psi_T$ has polynomially many non-zero elements in its Fourier transform. Observe that $\psi_T$ is the first row of the circulant $C_{T}$. The $j$th eigenvalue of $C_{T}$ is the $j$-th element of the Fourier transform of the first row ($\psi_T$) of $C_{T}$. This can be done through the sparse Fourier transform (SFT) algorithm in polynomial time \cite{Hassanieh2012a, Hassanieh2012b}. That is, the diagonal matrix $U(\Lambda_{C_{T}})$ is computable in polynomial time, when $T$ is frequency sparse.

Note that the unitary matrix $U(\Lambda_{C_{T}})$ is not necessarily Hermitian.  Simulating a non-Hermitian Hamiltonian can be done by Hermitian embedding as proposed by Jordan and Wocjan \cite{jordan2009}.  For completeness, we briefly describe the Jordan and Wocjan procedure below.
To start with, one embeds the unitary matrix $U(\Lambda_{C_{T}})$ in a Hermitian matrix,
\begin{equation}
H(U(\Lambda_{C_{T}})) = \begin{pmatrix} O & U(\Lambda_{C_{T}})  \\ {U(\Lambda_{C_{T}})}^{\dagger} & O \end{pmatrix}.
\end{equation}
Note that $H(U(\Lambda_{C_{T}}))$ is an involutory Hermitian matrix, i.e.
\begin{equation}
H(U(\Lambda_{C_{T}}))^2 = \begin{pmatrix} U(\Lambda_{C_{T}}) {U(\Lambda_{C_{T}})}^{\dagger} & O  \\ O & {U(\Lambda_{C_{T}})}^{\dagger} U(\Lambda_{C_{T}}) \end{pmatrix} = I.
\end{equation}
Also, the Euclidean norm of $H(U(\Lambda_{C_{T}}))$ is of unit value, namely $\left \| H(U(\Lambda_{C_{T}})) \right \| = 1 $.  It follows that,
\begin{equation}
e^{-iH(U(\Lambda_{C_{T}})) \theta} = \cos \theta I - i \sin \theta H(U(\Lambda_{C_{T}})),
\end{equation}
and we have
%
%
%
\begin{equation}
H(U(\Lambda_{C_{T}})) = i e^{-i \frac{\pi}{2} H(U(\Lambda_{C_{T}})) },
\end{equation}
as given by Jordan and Wocjan \cite{jordan2009}.

Since $U(\Lambda_{C_{T}})$ is row-computable and 2-sparse, it is an immediate observation that $H(U(\Lambda_{C_{T}}))$ is row-computable and 2-sparse as well. According to the sparse Hamiltonian lemma, $e^{-iH(U(\Lambda_{C_{T}})) \theta}$ (and therefore $H(U(\Lambda_{C_{T}}))$) is efficiently implementable \cite{jordan2009} \cite{aharanov2003}.

Let $\ket{\tilde{\psi}} = \ket{0} \ket{\psi}$ and we write ${U(\Lambda_{C_{T}})}^{\dagger}$ as,
\begin{equation}
{U(\Lambda_{C_{T}})}^{\dagger} =
 \begin{pmatrix}
  {{u}^{\dagger}}_{11} & {{u}^{\dagger}}_{12} \\
  {{u}^{\dagger}}_{21} & {{u}^{\dagger}}_{22} \\
 \end{pmatrix}.
\end{equation}
It can be seen that,
\begin{equation}
\begin{pmatrix} O & U(\Lambda_{C_{T}}) \\ {U(\Lambda_{C_{T}})}^{\dagger} & O \end{pmatrix}
\begin{pmatrix} 0 \\ 0 \\ F \tilde{\psi} \\ 0 \end{pmatrix}
=
 \begin{pmatrix}
   \frac{1}{k} \Lambda_{C_{T}} F \tilde{\psi}  \\
  \sqrt{I- \frac{1}{k^2}  \Lambda_{C_{T}}^{\dagger} \Lambda_{C_{T}} }  F  \tilde{\psi} \\
  0 \\
  0 \\
 \end{pmatrix}.
\end{equation}
Finally, we have
\begin{equation}
 \begin{pmatrix}
  F^{\dagger} &  &  &  \\
   &  F^{\dagger} & & \\
  & & F^{\dagger} & \\
  & & & F^{\dagger} \\
 \end{pmatrix}
\cdot
 \begin{pmatrix}
  O &  U(\Lambda_{C_{T}})  \\
   U(\Lambda_{C_{T}})^{\dagger} & O \\
 \end{pmatrix}
\cdot
 \begin{pmatrix}
  F &  &  &  \\
   &  F & & \\
  & & F & \\
  & & & F \\
 \end{pmatrix}
 \begin{pmatrix}
   0  \\
   0 \\
  \tilde{\psi} \\
  0 \\
 \end{pmatrix}
=
 \begin{pmatrix}
   \frac{1}{k}  F^{\dagger} \Lambda_{C_{T}} F \tilde{\psi}  \\
   F^{\dagger}  \sqrt{I- \frac{1}{k^2}  \Lambda_{C_{T}}^{\dagger} \Lambda_{C_{T}} }  F  \tilde{\psi} \\
  0 \\
  0 \\
 \end{pmatrix}.
\end{equation}
In Dirac notation, it can be expressed as follows.
\begin{equation}
	\label{eqn:dirac}
	\displaystyle
	\begin{array}{ll}
		  & \left( I_{4} \otimes F^{\dagger}_{2n} \right)  i e^{-i \frac{\pi}{2} H(U(\Lambda_{C_{T}}))}   \left( I_{4} \otimes F_{2n} \right) \ket{1} \ket{0} \ket{0} \ket{\psi}  \\
		  & \\
		= & \ket{0}  \left( \ket{0} \frac{1}{k} F^{\dagger}_{2n} \Lambda_{C_{T}} F_{2n} \ket{0}  \ket{\psi} + \ket{1} F^{\dagger}_{2n} \sqrt{I- \frac{1}{k^2} \Lambda_{C_{T}}^{\dagger} \Lambda_{C_{T}} } F_{2n}  \ket{0} \ket{\psi} \right) \\
		& \\
		= & \ket{0}  \left( \frac{1}{k} \ket{0} (\ket{0} T \ket{\psi} + \ket{1} B_{T} \ket{\psi}) + \ket{1} F^{\dagger}_{2n} \sqrt{I- \frac{1}{k^2} \Lambda_{C_{T}}^{\dagger} \Lambda_{C_{T}} } F_{2n}  \ket{0} \ket{\psi} \right) .
  	\end{array}
\end{equation}
Accordingly, we have a quantum circuit to implement $T \ket{\psi}$, as shown below.
 
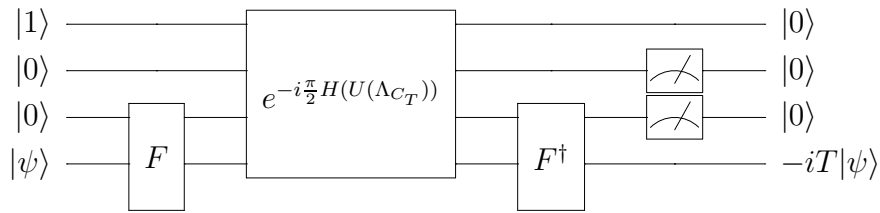
\[
 \Qcircuit @C=2.0em @R=0.1em @!R{
\lstick{\ket{1}}   & \qw     & \multigate{3}{e^{-i \frac{\pi}{2} H(U(\Lambda_{C_{T}}))}}   & \qw & \qw   &   \rstick{\ket{0}}  \qw \\
\lstick{\ket{0}}   & \qw     & \ghost{e^{-i \frac{\pi}{2} H(U(\Lambda_{C_{T}}))}}          & \qw   & \meter       & \rstick{\ket{0}} \qw \\
\lstick{\ket{0}}   & \multigate{2}{F}   & \ghost{e^{-i \frac{\pi}{2} H(U(\Lambda_{C_{T}}))}}   & \multigate{2}{F^{\dagger}}  & \meter   & \rstick{\ket{0}} \qw  \\
\lstick{\ket{\psi}}  & \ghost{F}    & \ghost{e^{-i \frac{\pi}{2} H(U(\Lambda_{C_{T}}))}}   & \ghost{F^{\dagger}}  & \qw      & \rstick{-i T \ket{\psi}} \qw \\
&              &              &       &          &     \\
 }
 \]

By sparse Hamiltonian simulation, $ e^{-i \frac{\pi}{2} H(U(\Lambda_{C_{T}}))}$ can be implemented efficiently, so $H(U(\Lambda_{C_{T}}))$, accordingly. Given the state $\ket{\psi}$ that we need to apply the Toeplitz $T$ on, we may append three qubits in the state $\ket{100}$ to $\ket{\psi}$. Then it will be followed by the sequence of operations $I_{4} \otimes F_{2n}$, $e^{-i \frac{\pi}{2} H(U(\Lambda_{C_{T}}))}$ and $I_{4} \otimes F^{\dagger}_{2n}$. Measurement of the first three qubits in the standard basis, conditioned on seeing $\ket{000}$ collapses the system to state $-i T\ket{\psi}$. (The additional $i$ is a global phase, that can be ignored.) It can be observed that the first qubit is already in the state $\ket{0}$, which makes its post-selection a deterministic operation. However, measurement of the second and the third qubits makes our algorithm probabilistic, as there is no guarantee they would be in the state $\ket{00}$. The probability of the measurement outcomes to be in the desired states is $\left \| \frac{1}{k} T \ket{\psi} \right \|^2$, which implies, whenever there is at least one entry in $T\ket{\psi}$ that is not exponentially small, the algorithm can be repeated to get the desired result efficiently.

\section{Applications and extensions}
From a large class of applications of Toeplitz matrices~\cite{michael1979, noschese2013}, we briefly describe a few tasks that can be done with our method, with an exponential saving, compared to the classical ways of performing them.

\subsection{Laplacians and banded Toeplitz matrices}
One significant and straightforward application is the calculation of Laplacians. Let us consider the second order Laplacian $L_2$, which is derivable from the second order central differences. The Laplacian $L_2$ is a banded Toeplitz matrix, which takes the following form:
\begin{equation}
	L_{2} = \left( \begin{array}{ccccccc}
                    2 & -1 & 0 & \cdots & 0 & 0 & 0 \\
                    -1 & 2 & -1 & \cdots & 0 & 0 & 0 \\
		    0 & -1 & 2 & \cdots & 0 & 0 & 0 \\
		    \vdots & \vdots & \vdots & \ddots & \vdots & \vdots & \vdots \\
		    0 & 0 & 0 & \cdots & 2 & -1 & 0 \\
		    0 & 0 & 0 & \cdots & -1 & 2 & -1 \\
		    0 & 0 & 0 & \cdots & 0 & -1 & 2 \\
                  \end{array}
                \right).
\end{equation}	
Consider a rotating system with cyclic symmetry (such as fans, compressors, or turbines \cite{Olson2014}) consisting of $n+2$ sectors, where the displacement of the  $i$th sector is denoted by $u_i$.  We can write down the discretised approximation of its acceleration, using second order central difference, as
\begin{equation}
\ddot{u_{i}} \approx \frac{u_{i+1}-2u_{i}+u_{i-1}}{h^2},
\end{equation}
where $h$ is the distance between two sectors.  Taking the system boundary conditions as $u_{0}= u_{n+1} = 0$, the acceleration vector can be expressed as $\ddot{u} =  - \frac{1}{{h^2}}  L_{2} u$. We can encode the displacement vector $u=\left( u_{\left( 0 \right)} , u_{\left( 1 \right)} , \cdots , u_{\left( n+1 \right)} \right)^{T}$ of the system in a quantum state with resource cost $log(n)$ using, for example, the Quantum Random Access Memory proposed by Giovannetti {\it et. al.}~\cite{Giovannetti2008}. Since $L_2$ is a sparse Toeplitz matrix, we can obtain the accelerations of all sectors in this system efficiently using the quantum circuit proposed above.

\subsection{Sparse circulant systems}
Our algorithm also provides an alternative way of solving sparse circulant systems efficiently. The inverse of the non-singular circulant
$C = circ(c_{1}, c_{2}, ... , c_{n})$ is given by,
\begin{equation}
C^{-1} = F^{\dagger} diag\left( \frac{1}{\lambda_1}, \frac{1}{\lambda_2}, \dots, \frac{1}{\lambda_n} \right) F.
\end{equation}
If $C$ is sparse, then the eigenvalue reciprocals in the above diagonal matrix can be computed efficiently; proving it to be a row-computable matrix. Following the above described steps, including the unitary dilation and Hermitian embedding, it can be readily seen that $C^{-1}$ is efficiently implementable. Recall that the HHL algorithm can also implement $C^{-1}$ efficiently, however the HHL circuit involves a phase estimation circuit, a number of Hadamard gates and controlled rotations, which are not present in our circuit.  The quantum circuit described in this paper serves as an alternative algorithm to solve a sparse circulant system, which is conceptually simpler and therefore may lead to a more efficient physical implementation than the HHL algorithm.

\subsection{Hankel matrices}
Recall that a Toeplitz matrix has constant elements in its diagonals. Contrastingly, a Hankel matrix has constant elements in its skew-diagonals. More precisely, a Hankel matrix is of the following form,
\begin{equation}
	H = \left( \begin{array}{ccccccc}
                    h_{-(n-1)} & h_{-(n-2)} & h_{-(n-3)} & \cdots & h_{-2} & h_{-1} & h_{0} \\
                    h_{-(n-2)} & h_{-(n-3)} & h_{-(n-4)} & \cdots & h_{-1} & h_{0} & h_{1} \\
		    h_{-(n-3)} & h_{-(n-4)} & h_{-(n-5)} & \cdots & h_{0} & h_{1} & h_{2} \\
		    \vdots & \vdots & \vdots & \ddots & \vdots & \vdots & \vdots \\
		    h_{-2} & h_{-1} & h_{0} & \cdots & h_{n-5} & h_{n-4} & h_{n-3} \\
		    h_{-1} & h_{0} & h_{1} & \cdots & h_{n-4} & h_{n-3} & h_{n-2} \\
		    h_{0} & h_{1} & h_{2} & \cdots & h_{n-3} & h_{n-2} & h_{n-1}
                  \end{array}
                \right).
\end{equation}
It is possible to permute a Hankel matrix into a Toeplitz matrix. Mathematically, this can be done by multiplying $H$ by the following permutation matrix.
\begin{equation}
P=\left( \begin{array}{ccccc}
                                        0 & 0 & \cdots & 0 & 1 \\
									    0 & 0 & \cdots & 1 & 0 \\
										\vdots & \vdots & \ddots & \vdots & \vdots \\
										0 & 1 & \cdots & 0 & 0 \\
										1 & 0 & \cdots & 0 & 0
                  \end{array}
                \right)
\end{equation}

Note that the matrix $P$ is efficiently implementable in a quantum circuit, as it is equal to the tensor product of Pauli x-operators. Let $T_{H}$ be the corresponding Toeplitz matrix; that is, $T_{H}=H P$. Also it follows that $H=T_{H} P$.

Consider the quantum state $\ket{\psi}$. Apply a Pauli x-gate to this state and make the state $P \ket{\psi}$. Now, we may implement $T_{H}$ on a quantum state $P\ket{\psi}$ as described in section 2. The respective measurement and post-selection gives an outcome proportional to  $T_{H}P\ket{\psi}$, which is equal to $H\ket{\psi}$. The runtime and the resource cost for the Hankel implementation is almost the same as for the Toeplitz implementation, since the Pauli x-gates are efficiently implementable in polynomial time and resource cost. Applications of Hankel matrices can be found in \cite{peller2012}.

\section{Discussion and conclusion}
We present quantum algorithms for implementing arbitrary row-sparse or frequency-sparse Toeplitz and Hankel matrices, a class of matrices which has a number of applications in different fields. The application of a classical sparse or frequency-sparse Toeplitz matrix is exponential, whereas the quantum algorithms presented in this paper can implement them in polynomial time; gaining exponential speedup over the classical procedures. It is noteworthy that the runtime of our algorithm is almost the same as the runtime of $ e^{-i \frac{\pi}{2} H(U(\Lambda_{C_{T}}))}$ for 2-sparse $H$. The two Fourier transforms has runtime $O(($log $n)^2)$, resulting in overall $O($polylog$(n))$ runtime. The probability of the measurement outcomes to be in the desired states is
$\left \| \frac{1}{k} T \ket{\psi} \right \|^2$. Following the sparse Hamiltonian lemma, our algorithms can implement any sparse or frequency-sparse Toeplitz or Hankel matrix $T$ on a state $\ket{\psi}$ to precision $\epsilon$, in runtime $O($ polylog $(n), \frac{1}{\epsilon}, \left \| \frac{1}{k} T \ket{\psi} \right \|^2)$.  This implementation is able to provide exponentially faster solutions to a variety of real-world problems.

\section{Acknowledgements}

We would like to thank Sisi Zhou, Thomas Loke, Josh Izaac and Lyle Noakes for valuable discussions on various aspects of quantum circuit design.  JBW also acknowledges discussions with Yogesh Joglekar on unitary dilation.

\bibliography{THreferences}

\end{document}